\documentclass[preprint,12pt]{elsarticle}

\usepackage[T1]{fontenc}
\usepackage[utf8]{inputenc}
\usepackage{amsmath,amssymb,amsfonts}
\usepackage{booktabs}
\usepackage{graphicx}
\usepackage{multirow}
\usepackage{url}
\usepackage{hyperref}
\usepackage{xcolor}
\usepackage{enumitem}
\usepackage{float}

\journal{Elsevier}

\begin{document}

\begin{frontmatter}

\title{Personalized electric vehicle energy consumption estimation framework that integrates driver behavior with map data}

\author[aff1]{Sreechakra Vasudeva Raju Rachavelpula\corref{cor1}}
\author[aff1]{Sangwhan Cha}
\cortext[cor1]{Corresponding author.}
\ead{smukherjee@my.harrisburgu.edu}
\ead{scha@harrisburgu.edu}

\address[aff1]{Department of Computer and Information Sciences, Harrisburg University of Science and Technology, Harrisburg, PA 17101, USA}

\begin{abstract}
This paper presents a personalized Battery Electric Vehicle (BEV) energy consumption estimation framework that integrates map-based contextual features with driver-specific velocity prediction and physics-based energy consumption modeling. The system combines route selection, detailed road feature processing, a rule-based reference velocity generator, a PID controller-based vehicle dynamics simulator, and a Bidirectional LSTM model trained to reproduce individual driving behavior. The predicted individual-specific velocity profiles are coupled with a quasi-steady backward energy consumption model to compute tractive power, regenerative braking, and State-of-Charge (SOC) evolution. Evaluation across urban, freeway, and hilly routes demonstrates that the proposed approach captures key driver behavioral patterns such as deceleration at intersections, speed-limit tracking, and road grade-dependent responses, while producing accurate power and SOC trajectories. The results highlight the effectiveness of combining learned driver behavior with map-based context and physics-based energy consumption modeling to produce accurate, personalized BEV SOC depletion profiles.
\end{abstract}


\end{frontmatter}

\section{Introduction}

The rapid transition from internal combustion engine vehicles to battery electric vehicles (BEVs) has intensified the need for accurate and reliable range estimation. Although BEVs offer substantial benefits in energy efficiency and emissions reduction, their adoption is hindered by persistent range anxiety, driven largely by inaccurate predictions of remaining driving range and state-of-charge (SOC). Existing range-estimation approaches typically rely on recent consumption averages or simplified driving assumptions, and therefore fail to account for critical contextual factors such as road geometry, traffic controls, and individual driving behavior. These omissions lead to significant deviations between predicted and actual energy use, particularly in urban environments with frequent stops, variable gradients, and heterogeneous driving styles.

Recent research highlights the importance of incorporating both map-based contextual information---such as slope, curvature, and speed limits---and driver-specific behavioral patterns, which can alter BEV energy consumption by 20--40\%. Advances in machine learning, especially recurrent neural networks and Long Short-Term Memory (LSTM) architectures, provide a promising foundation for modeling these temporal and contextual dependencies. When combined with detailed map data from platforms such as OpenStreetMap and Valhalla, LSTM-based models can forecast future velocity trajectories that reflect both learned past driver behavior and upcoming road conditions.

Building on these insights, this work develops a personalized, route context-aware BEV range estimation framework that integrates driver-specific velocity prediction with map-derived features and a physics-based energy consumption model. The goal is to produce more realistic velocity forecasts and ultimately deliver an improved energy consumption and SOC estimation system for electric vehicles.

\section{Literature Review}

Before we introduce our framework and system design, we examine the related literature published for the problems of vehicle speed prediction, driver behavior prediction, and map-context informed BEV energy consumption modeling. Vehicle speed prediction and energy consumption estimation have evolved from classical parametric models to sophisticated data-driven approaches that incorporate map information and driver behavioral patterns.

Early research relied on physics-based models such as the Intelligent Driver Model (IDM)~\cite{ref1}, which provided analytical formulations for car-following behavior and vehicle dynamics using simplified representations of traction, drag, and rolling resistance. However, these parametric models proved inadequate for real-world driving scenarios, failing to capture stochasticity and environmental variability in urban traffic conditions~\cite{ref2,ref3}. Markov chain extensions to the physics-based models bridged deterministic and stochastic approaches but suffered from accuracy degradation at longer prediction horizons due to first-order dependence assumptions~\cite{ref4}.

The availability of high-resolution telemetry data enabled machine learning paradigms, initially using Feedforward Neural Networks (FNN) and Support Vector Machines (SVM) for nonlinear mapping between driving parameters and velocities. Long Short-Term Memory (LSTM) networks became dominant for sequential prediction tasks, substantially outperforming shallow neural networks through selective information retention and sequence continuity modeling~\cite{ref2,ref7}. Bidirectional LSTMs further enhanced performance by processing sequences in both temporal directions, yielding smoother velocity profiles that mirror driver anticipation of upcoming road geometry~\cite{ref8,ref20}.

A critical advancement in velocity predictions involved incorporating map-derived features including road curvature, elevation, intersection density, and traffic signals. Studies demonstrated that integrating OpenStreetMap and elevation data substantially improved predictive accuracy by accounting for gravitational and frictional forces associated with road geometry~\cite{ref9,ref2}. Future road features proved essential for context-aware long-horizon forecasts, establishing map-aware prediction as state-of-the-art~\cite{ref10,ref11}.

Individual driving styles significantly impact energy consumption, with aggressive drivers reducing electric vehicle range by 20--40\% compared to conservative drivers~\cite{ref12,ref13}. Driver-specific modeling evolved from static behavioral clustering to temporal sequence modeling using recurrent networks, improving prediction robustness~\cite{ref14,ref15,ref16}. However, few works explicitly encode driver identity as a learnable feature, instead aggregating across multiple drivers and diluting individual distinctions.

Energy consumption and SOC estimation addresses range anxiety through physics-based backward vehicle models that decompose power demand into aerodynamic, inertial, and gravitational components~\cite{ref17,ref18}. Modern approaches integrate predictive velocity forecasting with energy management systems, demonstrating that accurate speed prediction directly improves energy allocation and range estimation~\cite{ref19}.

Despite significant advances, behavioral variability and route uncertainty remain major sources of estimation error in energy consumption. Current literature lacks comprehensive personalized frameworks that jointly encode driver-specific behavior and future road geometry for long-horizon prediction. This gap motivates the breadth of this research work.

\section{System Architecture}

\subsection{Motivation and Framework Overview}

\begin{figure}[htbp]
    \centering
     \includegraphics[width=\linewidth]{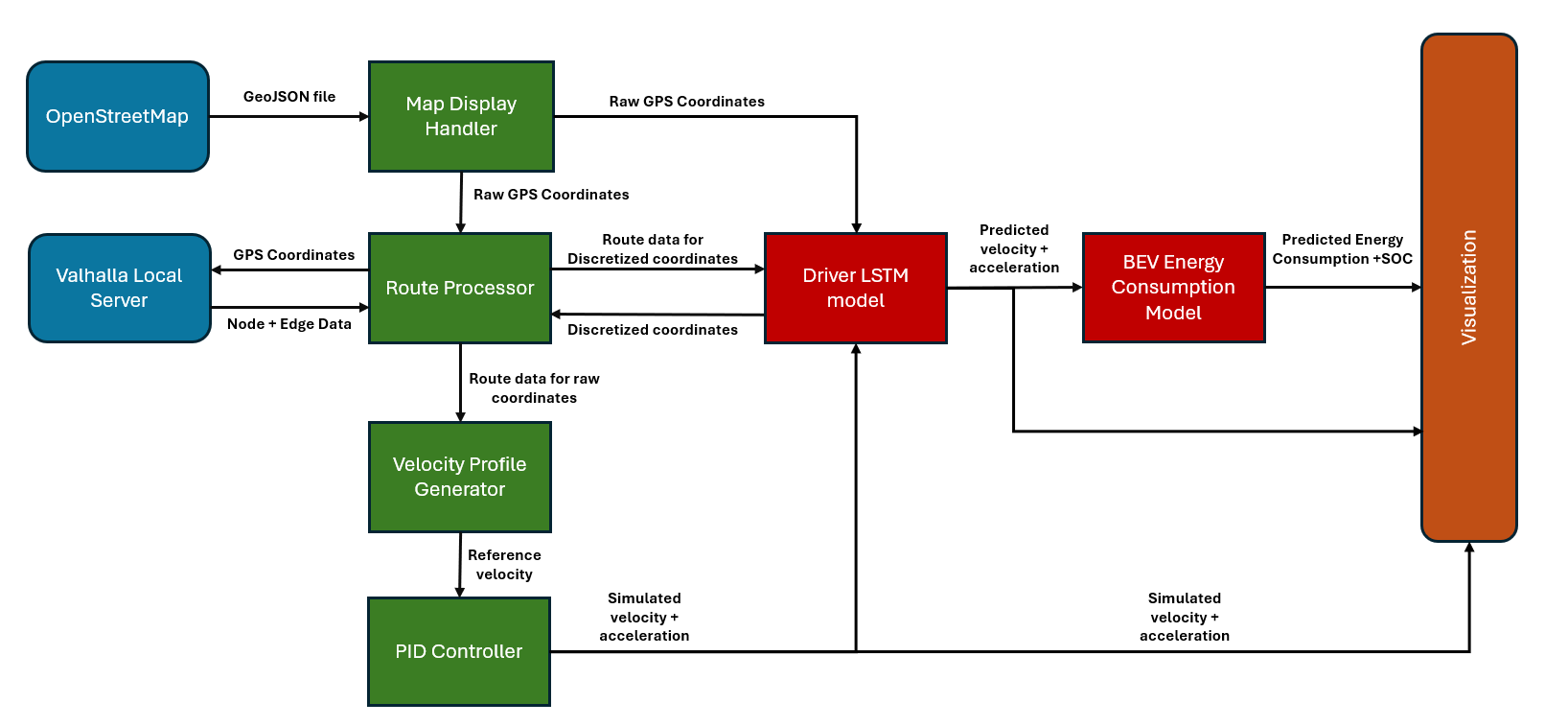}
    \caption{Overview of the proposed framework.}
    \label{fig:system_arch}
\end{figure}

The literature review identified three critical gaps in existing vehicle speed prediction and energy consumption estimation frameworks: (1) insufficient integration of driver-specific behavioral modeling with route data context, (2) limited capability for long-horizon personalized prediction, and (3) inadequate coupling between data-driven velocity forecasting and physics-based energy consumption models. To address these limitations, we propose a modular system architecture that unifies personalized driver modeling with comprehensive map feature integration and physics-based energy consumption estimation. The architecture comprises seven interconnected modules that produce personalized energy consumption estimates for a driver-selected route through a structured processing pipeline, illustrated in Fig.~\ref{fig:system_arch}.

\subsection{Module Descriptions and Data Flow}

\begin{enumerate}[label=\arabic*)]
  \item \textbf{Map Display Handler:} The pipeline begins with the Map Display Handler, which serves as the user interface for route selection. Implemented as a Python-Selenium wrapper around OpenStreetMap (OSM) controlled via ChromeDriver, this module enables users to specify start and destination points through address search or marker placement. The Selenium handler monitors URL changes and HTML elements to confirm successful route retrieval, prompting automatic download of the corresponding route geometry as a GeoJSON file containing latitude-longitude GPS coordinates. This geographic representation forms the main input for subsequent processing stages.

  \item \textbf{Route Processor:} The Route Processor transforms raw GPS coordinates into enriched road feature data by interfacing with the Valhalla routing engine through RESTful APIs. Using GPS coordinates from the GeoJSON file, the module sends HTTP POST requests to Valhalla's location and elevation API, retrieving node and edge-level attributes such as speed limits, curvature, slope, and intersection density. These API responses containing node and edge data are parsed into a structured \texttt{Pandas} DataFrame that serves as the comprehensive map feature dataset. Error-checking mechanisms handle incomplete or inconsistent API responses to maintain data integrity throughout the pipeline.

  \item \textbf{Velocity Profile Generator:} With structured map feature data available, the Velocity Profile Generator constructs an initial reference velocity trajectory. The module applies legal speed limits as upper bounds for straight segments, enforces reduced speeds for curves and roundabouts based on curvature severity, and sets zero velocity at traffic lights, stop signs, yield signs, and complex intersections. This rule-based profile provides a baseline velocity-distance curve that reflects fundamental map constraints without personalization, serving as the commanded input for vehicle dynamics simulation in the subsequent stage.

  \item \textbf{PID Controller:} To generate realistic driving data that captures vehicle dynamics such as acceleration limits, the PID Controller module simulates longitudinal motion by tracking the reference velocity through cascaded proportional-integral-derivative control. Through numerical integration of coupled ordinary differential equations, the controller produces smooth, jerk-limited, and acceleration-limited velocity profiles. This simulation generates time-series data including velocity, acceleration, jerk, and position that realistically represent how a vehicle would respond to route geometry and legal constraints. The resulting velocity profile serves as the ``past driving data'' required for personalized prediction in the Driver LSTM Module.

  \item \textbf{Driver LSTM Module:} The Driver LSTM Module represents the core contribution of the framework---a bidirectional LSTM network that performs personalized, map-aware velocity prediction. The module first discretizes the route into uniformly spaced 1-meter intervals and retrieves corresponding node attributes for these discretized coordinates using the Route Processor module, ensuring fine-grained spatial resolution for inference. Using the PID Controller generated velocity profile as past driving context and both past and future map features as environmental inputs, the pre-trained LSTM network conducts batch inference to predict driver-specific future velocities along the entire route.

  This prediction leverages the bidirectional LSTM architecture's ability to process temporal sequences in both forward and backward directions, effectively modeling how drivers anticipate upcoming road geometry when modulating speed. Post-processing using Butterworth and zero-phase filtering removes any high-frequency noise in the predictions while introducing zero phase lag. This results in a refined personalized velocity profile that is informed by both route context and individual driving style.

  \item \textbf{BEV Energy Consumption Model:} The predicted velocity profile feeds into the BEV Energy Consumption Model, which employs a physics-based backward model to estimate energy requirements. The model computes instantaneous acceleration from the velocity gradient, then calculates wheel tractive force by summing aerodynamic drag, rolling resistance, grade forces, and inertial forces. Mechanical power demand is determined from the tractive force and vehicle velocity, which is then used to obtain motor power and battery power consumption. Integrating these power profiles over time and distance yields total energy consumption and state-of-charge (SOC) depletion for the complete route.

  \item \textbf{Visualization Module:} The Visualization Module completes the pipeline by generating multi-panel graphical plots displaying predicted versus reference velocity profiles aligned with route distance, annotated with map features such as intersections, grade changes, and speed-limit zones. Additional visualizations show instantaneous power consumption at the wheels, motor, and battery subsystems, along with cumulative energy usage and SOC trajectories. These comparative plots enable rapid validation of prediction accuracy and facilitate interpretation of how map features and driving behavior jointly influence energy consumption.
\end{enumerate}

\subsection{Integrated Data Flow and Architecture Benefits}

The seven modules operate in sequence, with each stage editing or adding to the data stream:
\begin{enumerate}[label=\arabic*)]
  \item Route selection (Map Display Handler)
  \item Route feature extraction and structuring (Route Processor)
  \item Rule-based reference velocity profile generation (Velocity Profile Generator)
  \item Vehicle dynamics simulation with realistic constraints (PID Controller)
  \item Personalized, route context informed velocity and acceleration prediction (Driver LSTM Module)
  \item Physics-based energy consumption estimation (BEV Energy Consumption Model)
  \item Comprehensive visualization and validation (Visualization Module)
\end{enumerate}

This modular architecture directly addresses the identified research gaps by seamlessly integrating driver-specific behavioral modeling with comprehensive map context in a computationally efficient framework.

\section{Theoretical Background and Implementation Details}

The implementation details for the above described modules are briefly described in this section.

\subsection{Map Display Handler}

The Map Display Handler automates route acquisition from OpenStreetMap using Selenium-controlled browser interactions after the user selects a route. A ChromeDriver instance is launched and directed to \url{https://www.openstreetmap.org/directions}. The user specifies the start and end points, after which Selenium monitors URL changes and HTML page elements to detect successful route generation. Once detected, the module automatically triggers the ``Download route as GeoJSON'' option, producing a JSON file containing all GPS coordinates of the route for downstream processing.

\subsection{Route Processor}

The Route Processor extracts detailed node and edge-level features for each GPS coordinate using Valhalla's Location, Map Matching, and Elevation APIs. The coordinates of the selected route from the GeoJSON file are partitioned into batches and processed in parallel through a locally hosted Valhalla routing engine instance, with a public server at \url{https://valhalla1.openstreetmap.de} as fallback. API responses are merged into a unified dataset containing attributes such as speed limits, curvature, grade, elevation, heading, intersection type, and traffic control indicators. The resulting node-level attribute dataset forms the primary input to the Velocity Profile Generator and Driver LSTM modules.

\subsection{Velocity Profile Generator}

The Velocity Profile Generator applies a rule-based hierarchy to convert the road features into a reference velocity profile and speed-limit type that can be used by the PID Controller to generate a more physically accurate velocity profile. Each node is assigned a reference velocity and speed-limit type according to the following priority rules:
\begin{enumerate}[label=\arabic*)]
  \item \textbf{Traffic control elements (Type 3):} Presence of traffic signals, stop signs, or yield signs forces a stop (0 km/h).
  \item \textbf{Intersection entry (Type 3):} At 0\% edge position which also has a heading change of $> 60^{\circ}$, velocity is set to 0 km/h.
  \item \textbf{Intersection exit (Type 3):} At 100\% edge position which also has a heading change $> 60^{\circ}$, velocity is set to 0 km/h.
  \item \textbf{High curvature (Type 2):} If curvature exceeds a calibrated threshold, the reference velocity is limited to the average edge speed.
  \item \textbf{Default (Type 1):} Posted speed limit is used as the reference speed.
\end{enumerate}

\subsection{PID Controller}

\subsubsection{Theoretical Background}

The PID Controller simulates longitudinal vehicle dynamics through numerical integration of cascaded ordinary differential equations (ODEs) that model jerk- and acceleration-limited reference velocity tracking. The controller implements a three-state finite state machine---Standard Speed Limit Tracking, Curve Speed Tracking, and Stop Event Tracking. The states dynamically adjust the commanded velocity based on the look-ahead horizons computed from pre-calculated stopping distance lookup tables (LUTs).

\begin{figure}[htbp]
    \centering
     \includegraphics[width=\linewidth]{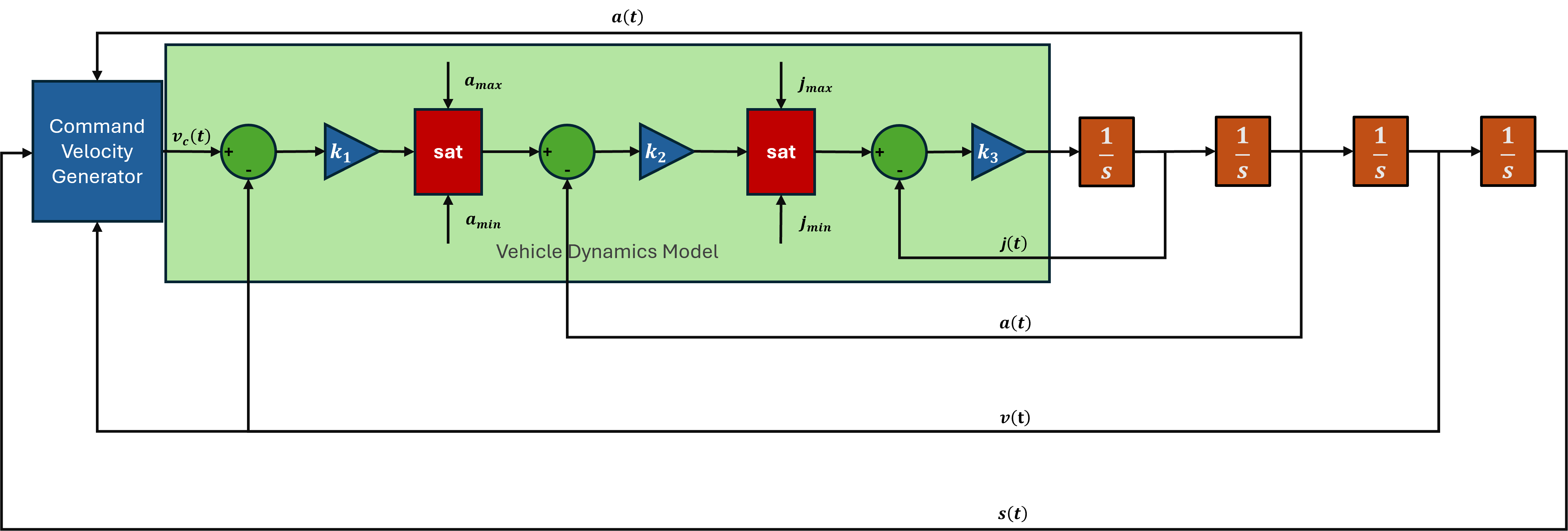}
    \caption{Block diagram representation of cascaded PID vehicle dynamics model.}
    \label{fig:VehicleDynModel}
\end{figure}

The vehicle dynamics are modeled as being governed by the cascaded PID system shown in Fig.~\ref{fig:VehicleDynModel} with saturation constraints defined by the following ODEs:
\begin{align}
\frac{ds}{dt} &= v, \\
\frac{dv}{dt} &= a, \\
\frac{da}{dt} &= j, \\
\frac{dj}{dt} &= k_3\Big(\operatorname{sat}\big[k_2\big(\operatorname{sat}[k_1(v_c-v); a_{\min}, a_{\max}] - a\big); j_{\min}, j_{\max}\big] - j\Big),
\end{align}
where $\operatorname{sat}(x;x_{\min},x_{\max})$ is the saturation function with $x_{\min}$ and $x_{\max}$ as the constraints for the state $x$. $\{s,v,a,j\}$ are the distance, velocity, acceleration, and jerk respectively. Tunable gains $k_1 < k_2 < k_3$ ensure stable tracking of the commanded speed $v_c$ with minimal overshoot. Default constraints are $a \in [-4,+2]$ m/s$^2$ and $j \in [-10,+10]$ m/s$^3$, for comfort-oriented driving.

\subsubsection{Implementation}

\begin{figure}[htbp]
    \centering
     \includegraphics[width=\linewidth]{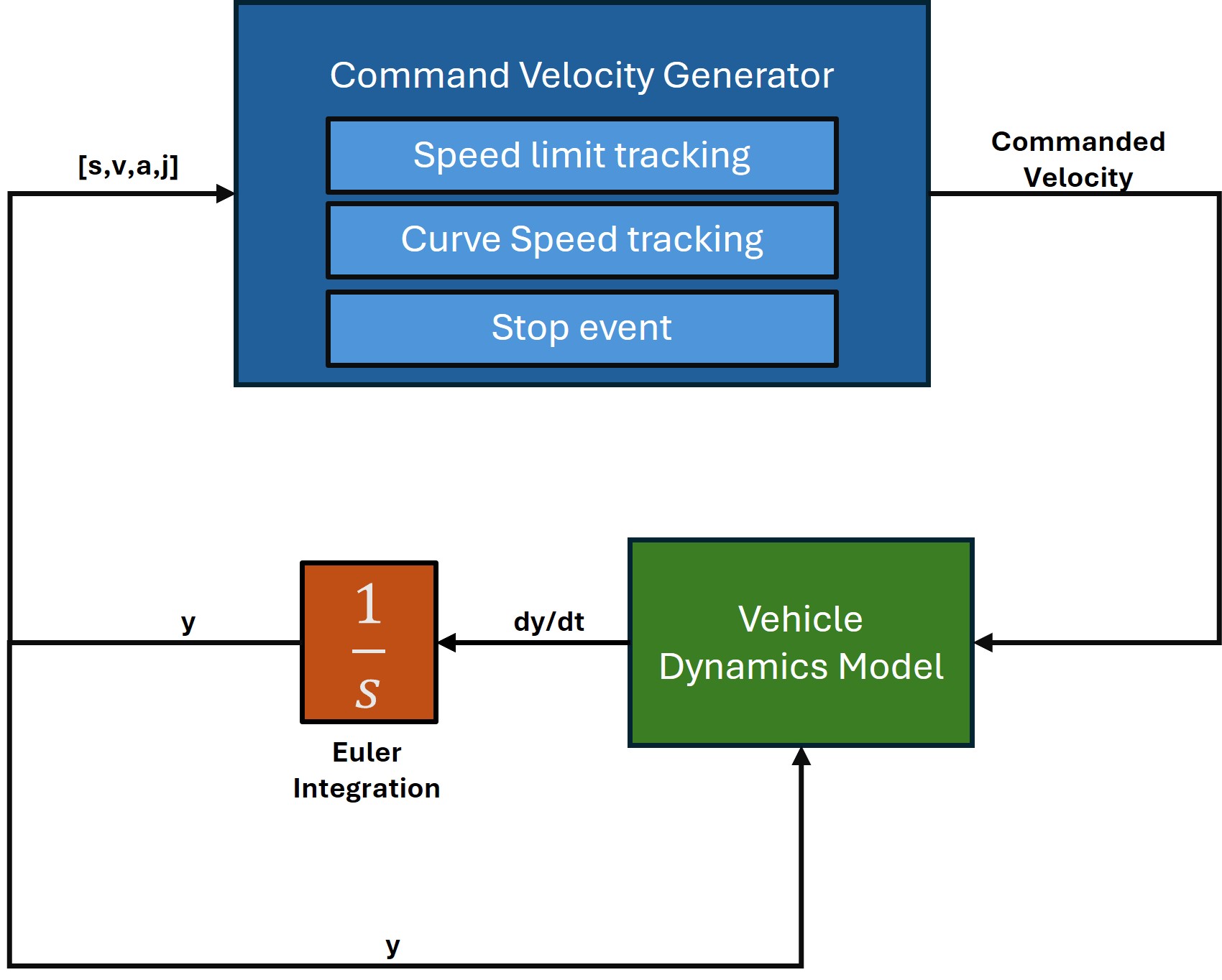}
    \caption{PID Controller module architecture showing Command Velocity Generator and Vehicle Dynamics Model interaction.}
    \label{fig:PIDController}
\end{figure}

The controller comprises two submodules as shown in Fig.~\ref{fig:PIDController}: the Command Velocity Generator produces the command velocity $v_c(t)$ through state-dependent logic, and the Vehicle Dynamics Model integrates the ODE system via Euler's method with a small timestep $h$ to minimize $O(h^2)$ truncation error.

\paragraph{Command Velocity Generator.} At each timestep, the module computes a look-ahead horizon,
\[
 s_h(t) = \mathrm{LUT}_{a_{\max}}(v(t)),
\]
using the worst-case stopping distances from a 1-D pre-computed LUT for initial acceleration, $a_0 = 2$ m/s$^2$. All speed limit change events within $s_h(t)$ are identified, and required deceleration distance $s_r(t)$ is computed as:
\begin{equation}
 s_r(t)=
 \begin{cases}
  \mathrm{LUT}_{a_{\max}}(v(t)), & \text{Type 3 (Stop)}, \\
  \mathrm{LUT}_{a_{\max}}(v(t))\,\dfrac{v(t)-v_f(t)}{v(t)}, & \text{Type 2 (Curve)}.
 \end{cases}
\end{equation}
where $v_f(t)$ is the reference velocity at the curve event.

State transitions occur when distance to event, $s_f(t)-s(t)$, drops below $s_r(t)$. Here $s_f(t)$ is the absolute distance to the future event. In Stop Event Tracking, the command velocity is set to $v_c(t)=0$ until the vehicle halts at or before $s_f(t)$. In Curve Speed Tracking, commanded velocity smoothly decelerates by setting:
\begin{equation}
 v_c(t)=v_f(t)-\left|v_f(t)\tanh\left(\frac{k\,s_r(t)}{s_f(t)-s(t)}\right)\right|,
\end{equation}
ensuring curve speed attainment before event onset.

\paragraph{Integration.} Euler's method advances the state vector $y=\{s,v,a,j\}^{T}$ via
\[
 y(t+h)=y(t)+f(t,y)h.
\]
The resulting velocity, acceleration, and jerk profiles provide realistic past driving sequences for the Driver LSTM Module's batch inferencing.

\subsection{Driver LSTM Module}

\subsubsection{Theoretical Background}

The Driver LSTM Module employs a sequence-to-sequence Bidirectional LSTM-based hybrid encoder-decoder architecture similar to the one used by Paparusso et al.~\cite{ref8} to capture temporal dependencies between past driving behavior and future road geometry for future vehicle state prediction. The framework addresses the inherently sequential, nonlinear nature of driver-vehicle dynamics conditioned by both historical behavior and upcoming road features.

\paragraph{Problem Formulation.} Given position index $k$, past window size $s_P$, and prediction horizon $s_F$, the model approximates:
\begin{equation}
 f_\theta : \Big(X^{vP}_{(k-s_P:k)}, X^{rP}_{(k-s_P:k)}, X^{rF}_{(k+1:k+s_F)}\Big) \mapsto Y^{vF}_{(k+1:k+s_F)}
\end{equation}
where $X^{vP} \in \mathbb{R}^{(s_P+1)\times n_v}$ contains past vehicle states such as velocity and acceleration, $X^{rP} \in \mathbb{R}^{(s_P+1)\times n_r}$ and $X^{rF} \in \mathbb{R}^{s_F\times n_r}$ contain past and future road features such as elevation, heading, speed limit, curvature, traffic control flags, and edge classification, respectively, and $Y^{vF} \in \mathbb{R}^{s_F\times q_v}$ represents predicted future vehicle states. Here $n_v$ is the number of input vehicle states, $n_r$ is the number of input road features, and $q_v \le n_v$ is the number of predicted vehicle states.

\subsubsection{Architecture and Implementation}

\begin{figure}[htbp]
    \centering
     \includegraphics[width=\linewidth]{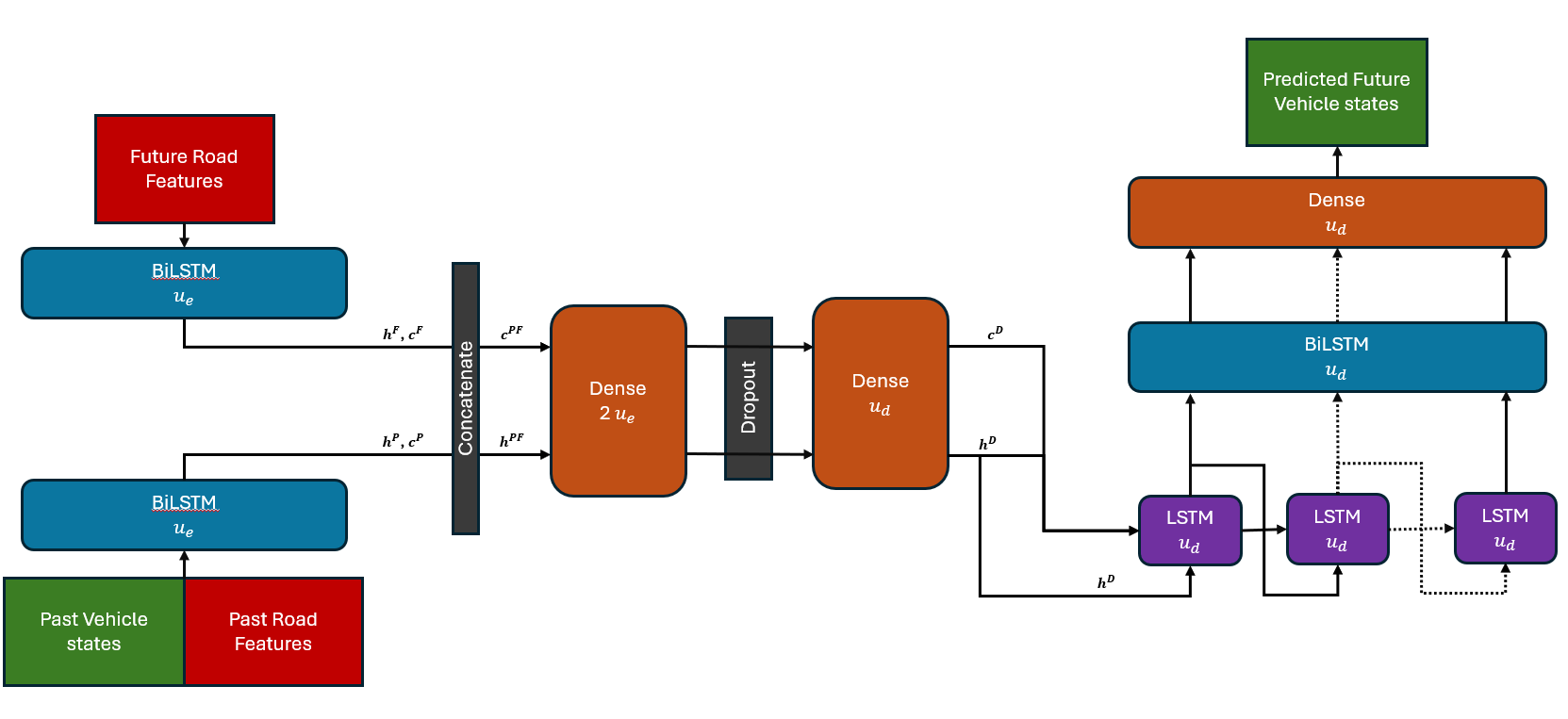}
    \caption{Overview of Driver LSTM Module encoder-decoder architecture.}
    \label{fig:DriverLSTM_arch}
\end{figure}

We propose the hybrid encoder-decoder architecture shown in Fig.~\ref{fig:DriverLSTM_arch} which comprises two encoders that project the past driving and future road context into latent states which are then used by a recursive decoder for predicting future vehicle states:
\begin{enumerate}[label=\arabic*)]
  \item \textbf{Past Driving Encoder:} Concatenated $X^{vP}$ and $X^{rP}$ are encoded via a BiLSTM layer ($u_e = 32$ units). Forward and backward final hidden states $(h^P_F,h^P_B)$ and cell states $(c^P_F,c^P_B)$ are concatenated to form unified past state representations $h^P$ and $c^P$.

  \item \textbf{Future Road Encoder:} $X^{rF}$ is encoded through a separate BiLSTM layer (of hidden size $u_e = 32$ units), producing similarly concatenated forward and backward hidden states $h^F$ and cell states $c^F$ which represent the future road context.

  \item \textbf{State Fusion:} Concatenated past and future hidden states $h^{PF}=[h^P,h^F]$ and cell states $c^{PF}=[c^P,c^F]$ are projected through two dense layers with $2u_e$ and $u_d$ neurons and a dropout ($r=0.25$) for regularization, yielding state vectors $h^D$ and $c^D$.

  \item \textbf{Recursive Decoder:} A unidirectional LSTM ($u_d = 64$ units) initialized with states $h^D$ and $c^D$ and the first cell fed with $h^D$ as input generates sequences autoregressively. The output sequence feeds into a BiLSTM layer ($u_d = 64$ units) followed by a time-distributed dense layer producing the sequence of future vehicle state predictions, $\hat{Y}^{vF}$.
\end{enumerate}

Input sequences use $s_P = 50$ past steps and $s_F = 100$ future prediction steps. Input features include the vehicle states longitudinal velocity and acceleration and road features elevation, heading, speed limit, average edge speed, curvature, traffic signals, stop/yield signs, roundabouts, edge classification, link flags, and engineered start-stop indicators. Vehicle speed is the predicted future variable. Continuous features are scaled via SciPy's \texttt{StandardScaler} while categorical features are unscaled.

\paragraph{Training.} The IEEE Vehicle Speed Dataset~\cite{ref21} containing 5973 rides (9049.3 km) at 1-meter spatial resolution from Czech Republic driving data was used for training the Driver LSTM model. Due to inconsistencies in the categorical road features of the original dataset, additional road features were generated using the Route Processor module and were appended to the original dataset. This amended dataset was used in training. Training employed Adam optimizer with learning rate 0.0001, mean-squared error loss function, 70:15:15 train/validation/test split, 30 epochs with early stopping, and sequence generation using stride length of 10 steps with padding. Multiple retraining iterations with data randomization were done to minimize mean absolute error.

\paragraph{Inference.} The selected route from the Map Display Handler is discretized to 1-meter intervals and the road features are retrieved for these discretized coordinates using Route Processor. The PID Controller generated velocity and acceleration profiles are then interpolated onto these discretized coordinates, forming the past vehicle states. These data points are subsequently processed, using a stride of 10 steps, into batches of sequences containing past vehicle states $X^{vP}$, past road-context features $X^{rP}$, and future road-context features $X^{rF}$. These batches of past and future feature matrices are fed into the loaded best LSTM model to predict the future velocity profile $\hat{Y}^{vF}$, from which only the first 10 predicted steps are retained for each inference window. The predicted velocity segments are then post-processed using a first-order Butterworth low-pass filter followed by a zero-phase forward-backward filter to remove high-frequency noise without phase distortion. The resulting personalized velocity profile feeds the BEV Energy Consumption Model for power, energy consumption, and SOC depletion estimation.

\subsection{BEV Energy Consumption Model}

\subsubsection{Theoretical Background}

BEV energy consumption and state of charge (SOC) estimation employ a quasi-steady, backward physics-based model following the VT-CPEM framework~\cite{ref18}. The model computes the instantaneous energy consumption using second-by-second vehicle speed, acceleration, and roadway grade data as input variables, enabling high-resolution estimation of tractive power, motor power, battery power, and SOC evolution along the route.

\paragraph{Tractive Power.} Total tractive force comprises inertial, grade, rolling resistance, and aerodynamic components:
\begin{equation}
 F_{\text{total}}(t)=ma(t)+mg\sin(\theta(t))+\frac{1}{2}\rho A_f C_D v(t)^2 + mg\cos(\theta(t))\big(C_2+C_1v(t)\big)\frac{C_r}{1000}.
\end{equation}
From the tractive force, the tractive power at the wheels is calculated as:
\begin{equation}
 P_{\text{wheels}}(t)=F_{\text{total}}(t)\,v(t)
\end{equation}
where $m$ is vehicle mass, $a$ acceleration, $v$ velocity, $\theta$ road grade angle, $C_r$ rolling resistance coefficient, $C_1,C_2$ empirical constants, $\rho$ air density, $A_f$ frontal area, and $C_D$ drag coefficient. Positive power indicates traction; negative indicates braking.

\paragraph{Motor Power.} During traction ($P_{\text{wheels}}\ge 0$), the motor supplies power to the wheels:
\begin{equation}
 P_{\text{motor}}(t)=\frac{P_{\text{wheels}}(t)}{\eta_{\text{driveline}}\,\eta_{\text{motor}}}.
\end{equation}
During braking ($P_{\text{wheels}}<0$), the motor operates in generator mode. The regenerative braking efficiency $\eta_{rb}$ is modeled as an exponential function of deceleration:
\begin{equation}
 \eta_{rb}(t)=
 \begin{cases}
  e^{-\left(\alpha/|a(t)|\right)}, & a(t)<0,\\
  0, & a(t)\ge 0,
 \end{cases}
 \qquad 0\le \eta_{rb}(t)\le \eta_{rb,\max}
\end{equation}
and
\begin{equation}
 P_{\text{motor}}(t)=P_{\text{wheels}}(t)\,\eta_{rb}(t)\,\eta_{\text{driveline}}\,\eta_{\text{motor}}.
\end{equation}
where $\eta_{\text{driveline}}$ and $\eta_{\text{motor}}$ are the driveline and motor efficiencies respectively.

\paragraph{Battery Power.} The net battery power accounts for motor power, auxiliary loads $P_{aux}$ (e.g., HVAC, electronics), and battery efficiency $\eta_{batt}$:
\begin{equation}
 P_{\text{batt}}(t)=
 \begin{cases}
  \dfrac{P_{\text{motor}}(t)+P_{aux}}{\eta_{batt}}, & \text{traction},\\
  \big(P_{\text{motor}}(t)+P_{aux}\big)\eta_{batt}, & \text{regeneration}.
 \end{cases}
\end{equation}
The sign convention ensures that positive battery power corresponds to discharge and negative power corresponds to charging.

\paragraph{Energy Consumption.} The energy consumption over a route segment is computed by integrating the instantaneous battery power over time. The final energy consumption per kilometer in Wh/km is computed as:
\begin{equation}
 EC=\frac{1}{3600\,d}\int P_{\text{batt}}(t)\,dt
\end{equation}
where $d$ is total distance (km).

\paragraph{SOC Dynamics.} Battery state of charge evolves according to:
\begin{align}
 \Delta SOC(t) &= \frac{P_{\text{batt}}(t)}{3600\,C_W}, \\
 SOC(t) &= SOC(t-1)-\Delta SOC(t),
\end{align}
where $C_W$ is battery capacity (Wh). SOC is constrained to the operating range $[20\%,95\%]$ for battery health preservation.

\subsubsection{Implementation}

The Driver LSTM model-predicted velocity profile feeds directly into the BEV Energy consumption model, propagating driver-specific speed and acceleration transients through tractive power, regenerative braking efficiency, and SOC calculations. This sensitivity to behavioral dynamics enables personalized range estimation reflecting both route geometry and individual driving style.

\section{Testing and Results}

To evaluate end-to-end system performance, the pipeline was tested on three representative route types designed to capture diverse driving conditions: urban route, freeway route, and hilly terrain route.

\subsection{Urban Route}

\begin{figure}[htbp]
    \centering
     \includegraphics[width=\linewidth]{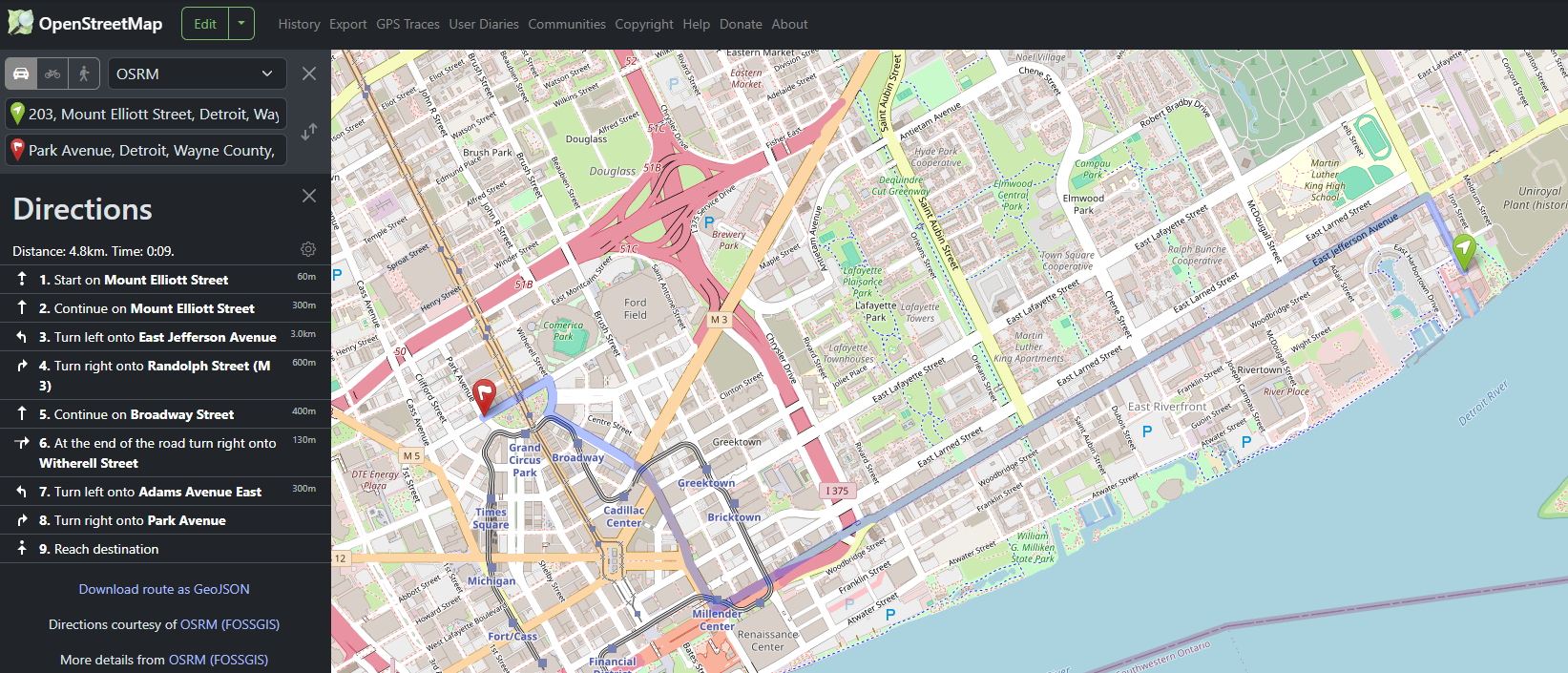}
    \caption{Map view of the chosen urban route.}
    \label{fig:urban_drive_route}
\end{figure}

\begin{figure}[htbp]
    \centering
     \includegraphics[width=\linewidth]{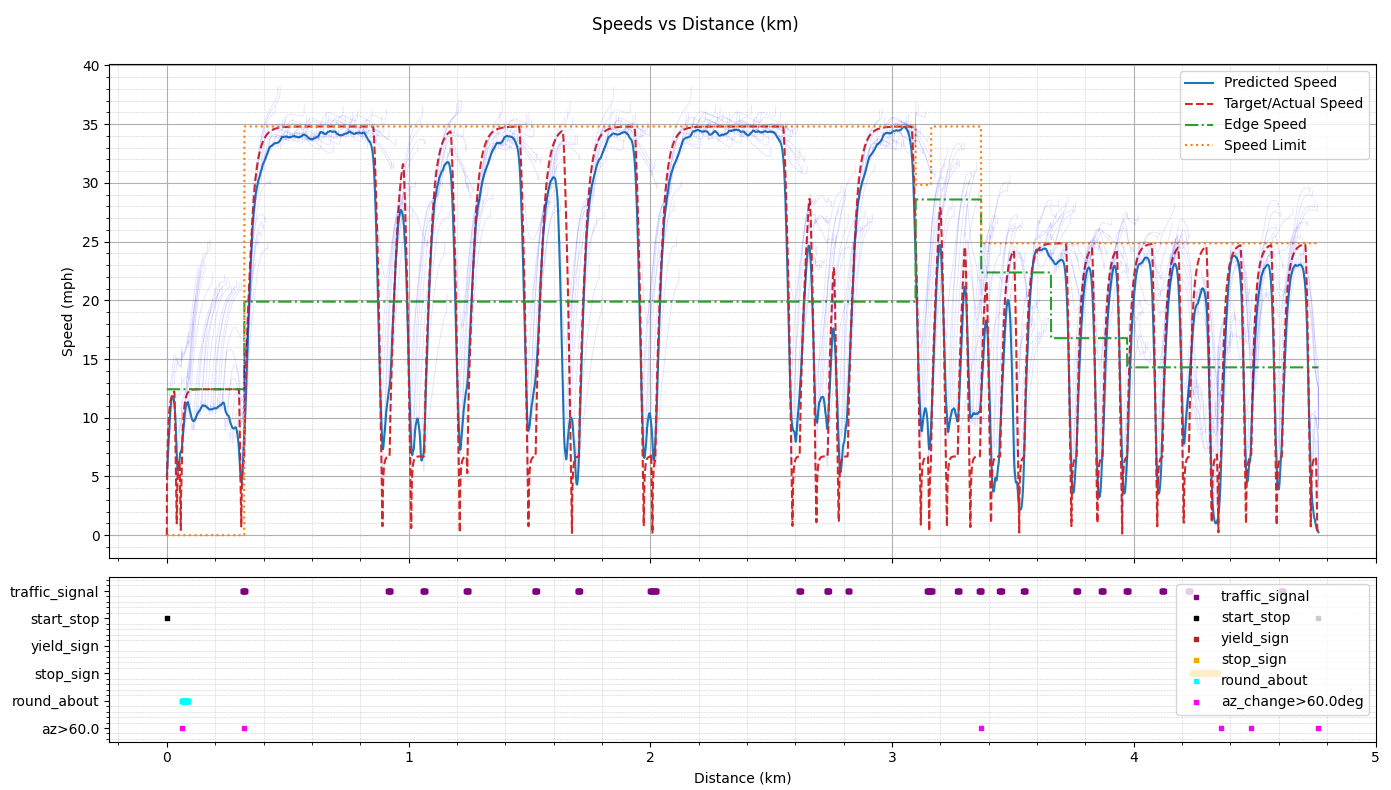}
    \caption{Predicted velocity for the chosen urban route.}
    \label{fig:urban_drive_vel}
\end{figure}

Urban driving exhibits repeated acceleration and deceleration due to traffic control devices and intersection density. On the Detroit downtown route in Fig.~\ref{fig:urban_drive_route}, the Driver LSTM-predicted velocity of Fig.~\ref{fig:urban_drive_vel} captures the deceleration trends near intersections but does not always reach a full stop, typically only dropping below 5 mph for each intersection. This behavior is likely due to the limited future context (100 m) chosen in the LSTM architecture, which may be insufficient to perfectly model full vehicle stops.

Additionally, the ground-truth dataset used for training does not always contain zero-speed samples for traffic signals or yield signs. As a result, the LSTM model appears to have learned that a full stop is not always required---an interpretation that aligns with real-world driving, where green traffic lights or yield signs during low traffic conditions often eliminate the need to stop. Nevertheless, the model correctly predicts deceleration behavior near traffic signals and at intersections with large heading changes (i.e., left or right turns).

\begin{figure}[htbp]
    \centering
     \includegraphics[width=\linewidth]{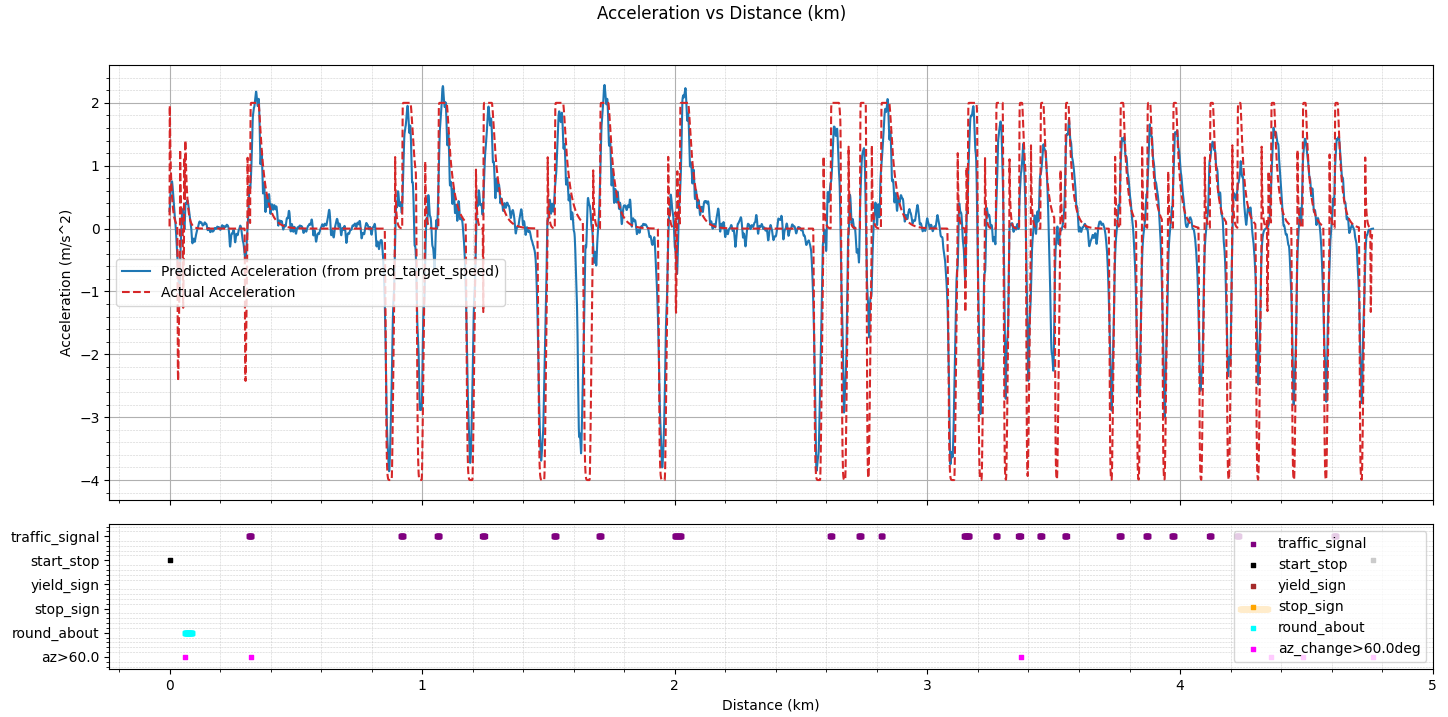}
    \caption{Predicted acceleration for the chosen urban route.}
    \label{fig:urban_drive_acc}
\end{figure}

The acceleration profile in Fig.~\ref{fig:urban_drive_acc} further illustrates the differences between the simulated and predicted profiles. The simulated acceleration from the PID Controller saturates at the predefined maximum and minimum limits, applying maximum acceleration during speed increases and maximum deceleration during braking. The driver-specific predicted acceleration, however, does not exhibit such saturation and instead reflects smoother, more realistic driving behavior. Both the predicted velocity and acceleration contain natural variations, unlike the smooth simulated profiles, capturing the variability present in real-world driving and the patterns learned from the training data.

\begin{figure}[htbp]
    \centering
     \includegraphics[width=\linewidth]{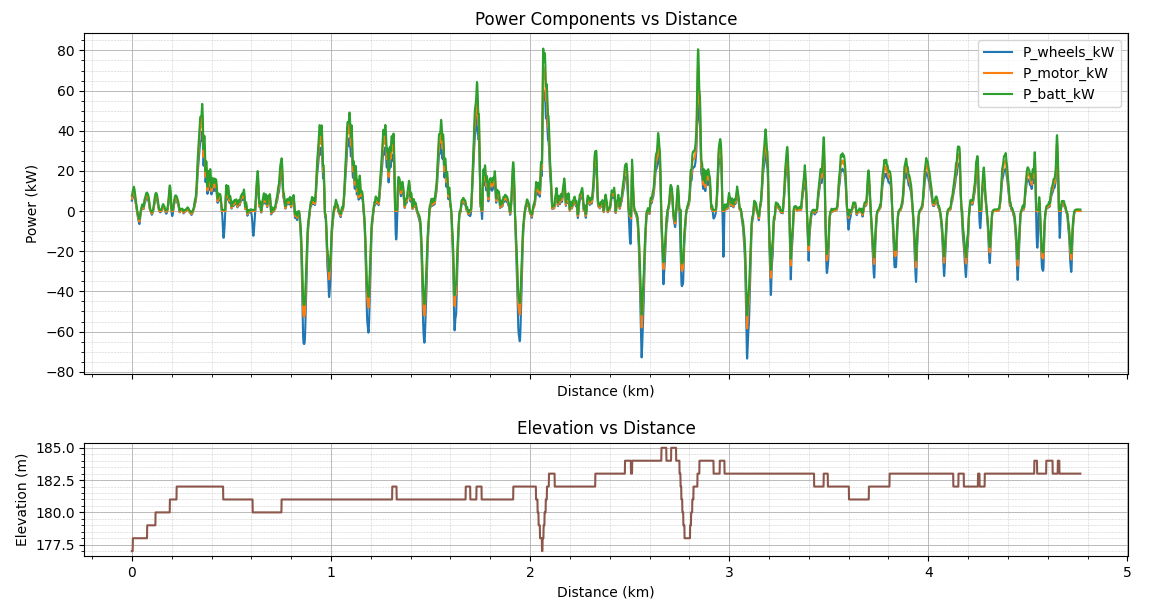}
    \caption{Predicted power demand for the chosen urban route.}
    \label{fig:urban_drive_power}
\end{figure}

\begin{figure}[htbp]
    \centering
     \includegraphics[width=\linewidth]{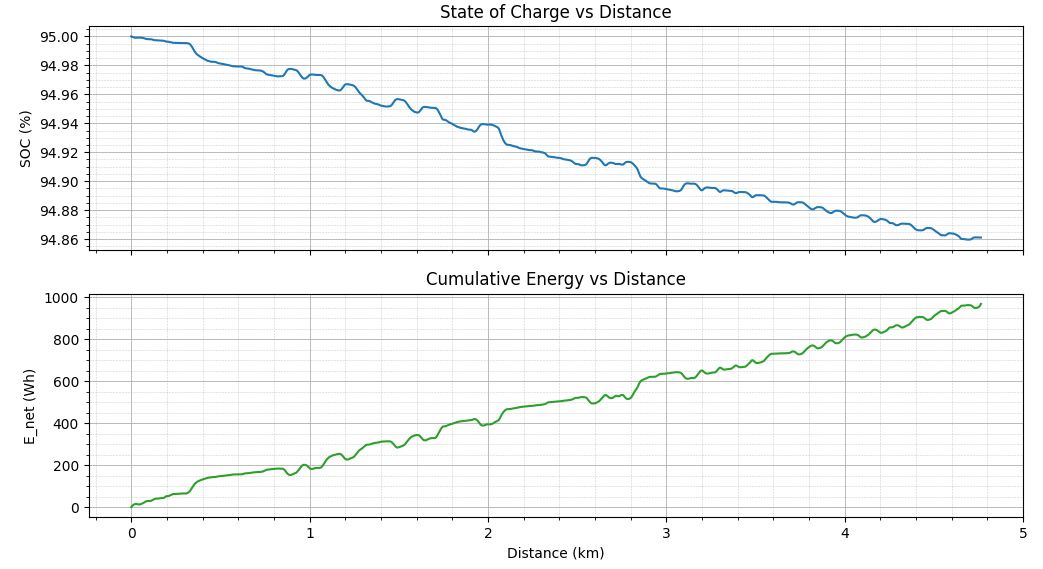}
    \caption{Predicted energy consumption for the chosen urban route.}
    \label{fig:urban_drive_soc}
\end{figure}

Power and SOC trajectories in Fig.~\ref{fig:urban_drive_power} and Fig.~\ref{fig:urban_drive_soc} also follow expected patterns: positive acceleration yields positive power demand, while braking events produce regenerative power and corresponding SOC recovery.



\subsection{Freeway Route}

\begin{figure}[htbp]
    \centering
     \includegraphics[width=\linewidth]{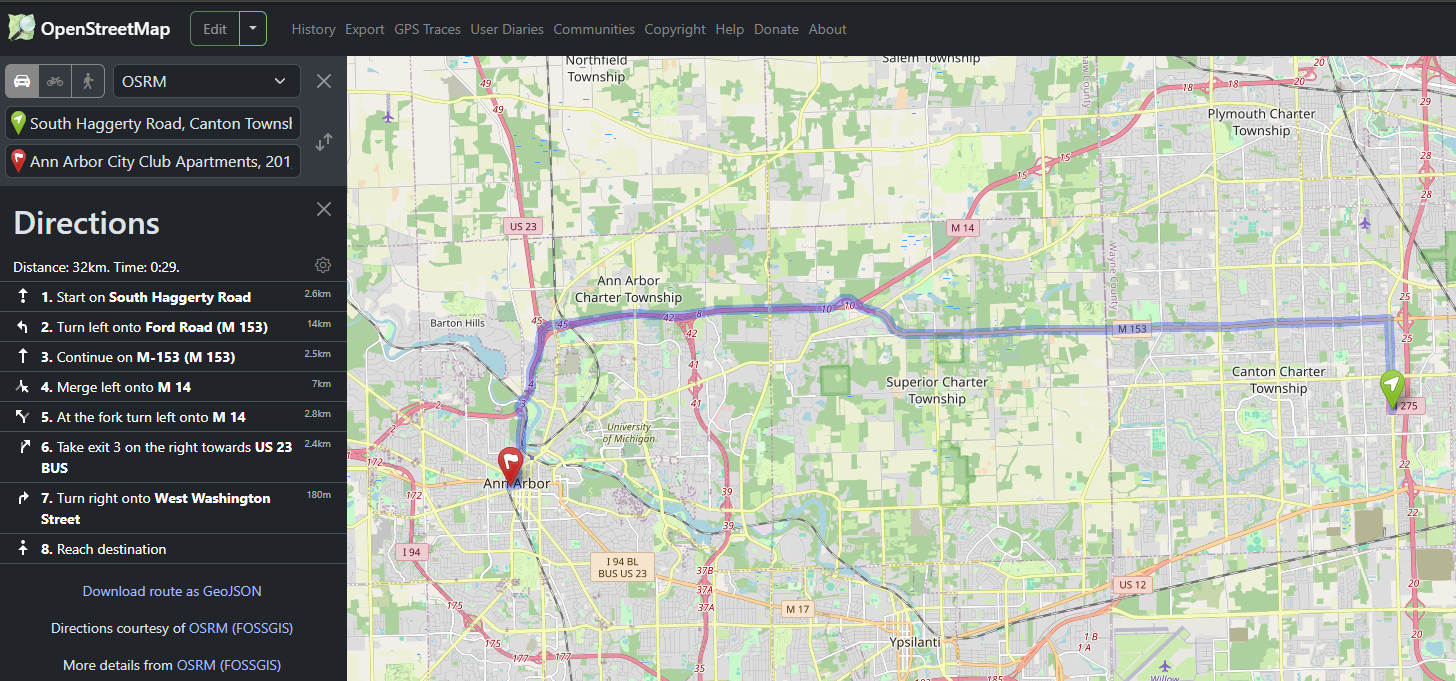}
    \caption{Map view of the chosen freeway route.}
    \label{fig:freeway_urban_route}
\end{figure}

\begin{figure}[htbp]
    \centering
     \includegraphics[width=\linewidth]{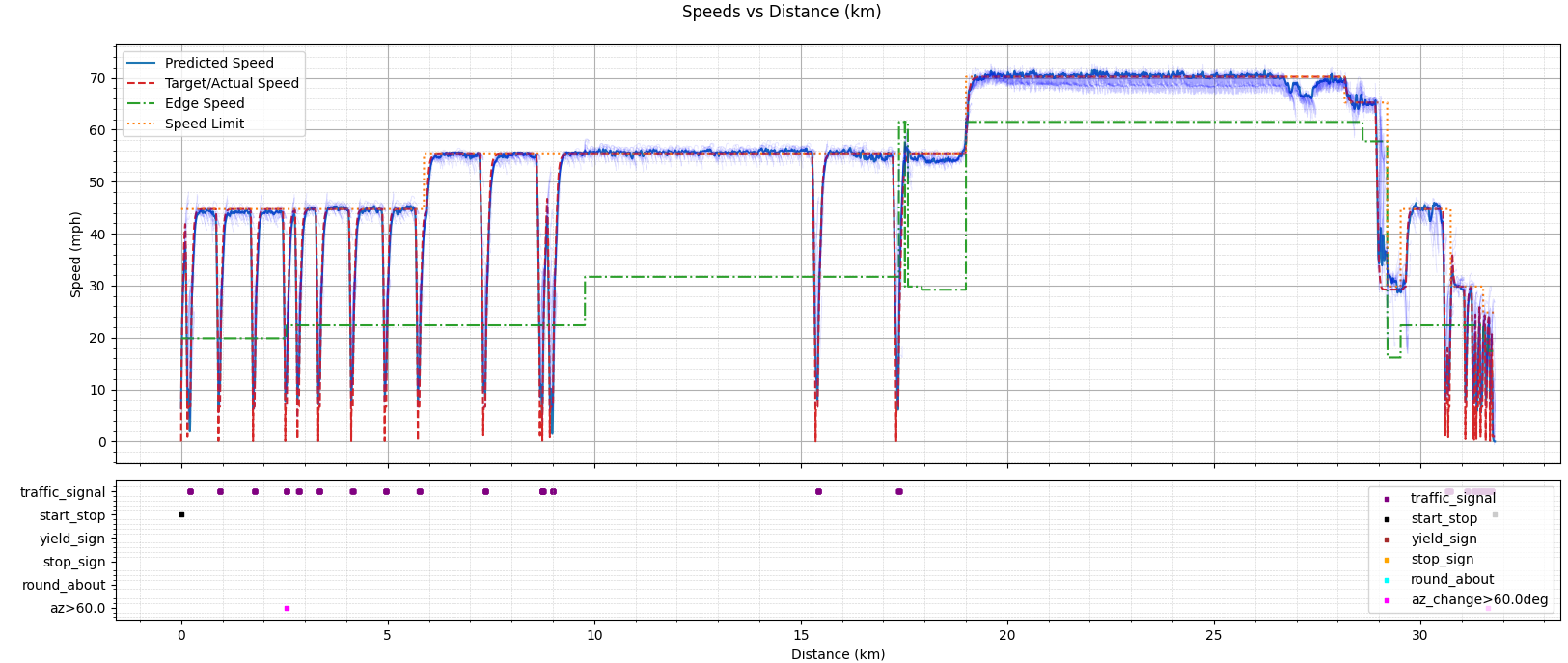}
    \caption{Predicted velocity for the chosen freeway route.}
    \label{fig:freeway_urban_vel}
\end{figure}

\begin{figure}[htbp]
    \centering
     \includegraphics[width=\linewidth]{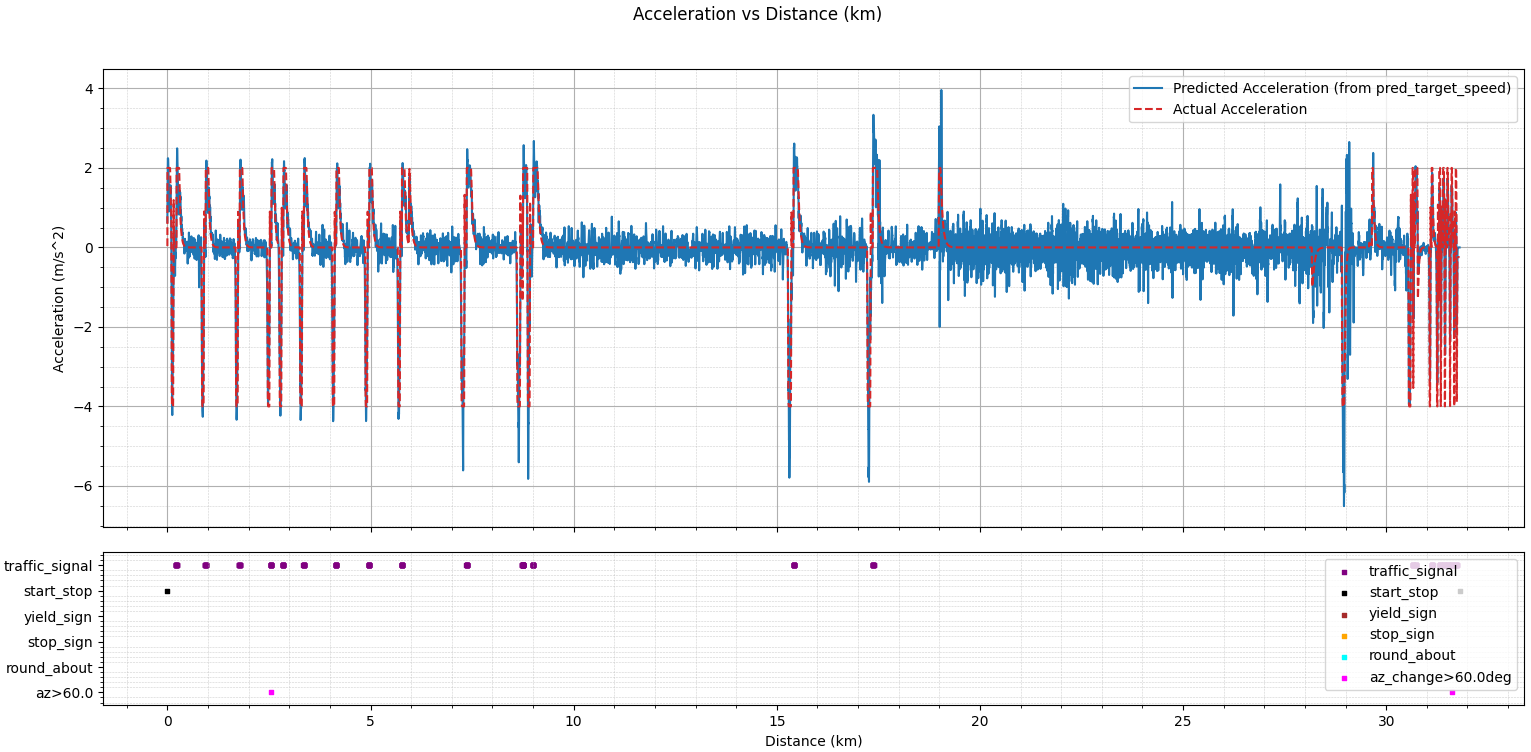}
    \caption{Predicted acceleration for the chosen freeway route.}
    \label{fig:freeway_urban_acc}
\end{figure}

The freeway route between Canton, MI and Ann Arbor, MI used for testing shown in Fig.~\ref{fig:freeway_urban_route} features long segments of steady cruising. The LSTM-predicted velocity in Fig.~\ref{fig:freeway_urban_vel} oscillates around the legal speed limit, reflecting typical driver micro-adjustments, whereas the PID reference speed used in inference remains smooth. These oscillations also appear in the predicted acceleration in Fig.~\ref{fig:freeway_urban_acc}, which reflects the driver's personal driving style.

\begin{figure}[htbp]
    \centering
     \includegraphics[width=\linewidth]{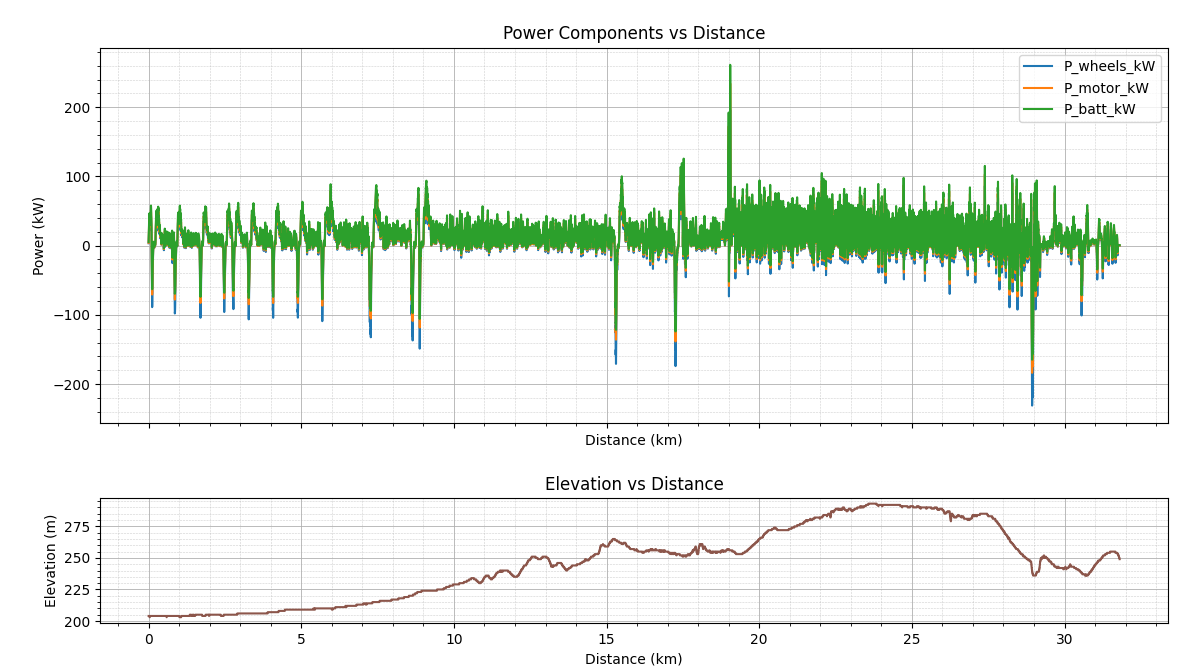}
    \caption{Predicted power demand for the chosen Freeway Route.}
    \label{fig:freeway_urban_power}
\end{figure}

\begin{figure}[htbp]
    \centering
     \includegraphics[width=\linewidth]{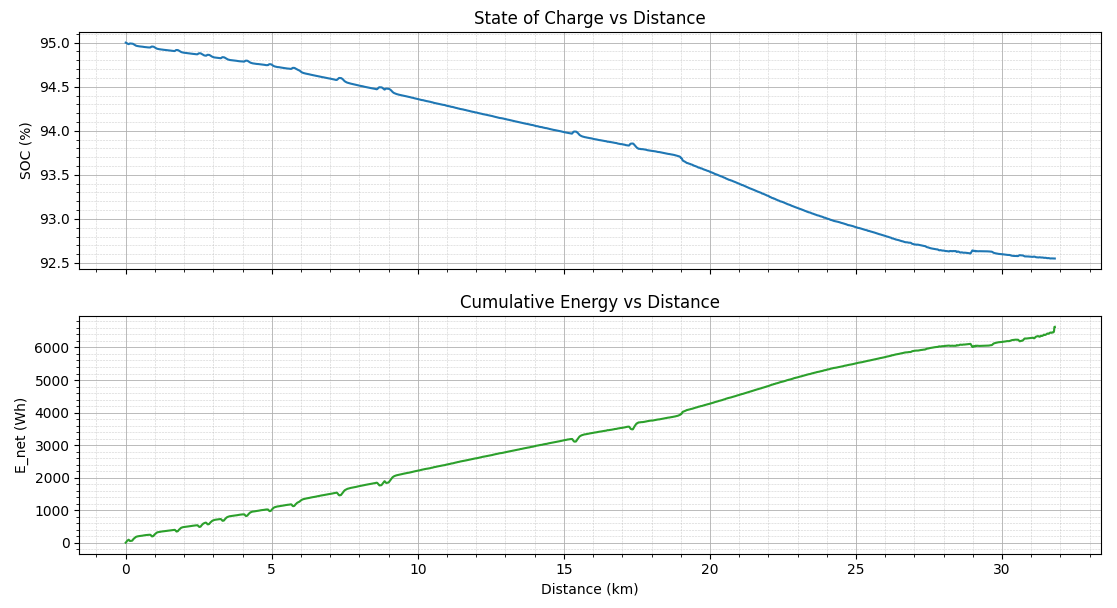}
    \caption{Predicted energy Consumption for the chosen Freeway Route.}
    \label{fig:freeway_urban_soc}
\end{figure}

Power and SOC profiles in Fig.~\ref{fig:freeway_urban_power} and Fig.~\ref{fig:freeway_urban_soc} show near-zero mean power fluctuations and almost linear SOC decline in the freeway sections, consistent with low-variability freeway driving.

\subsection{Hilly Terrain Route}

\begin{figure}[htbp]
    \centering
     \includegraphics[width=\linewidth]{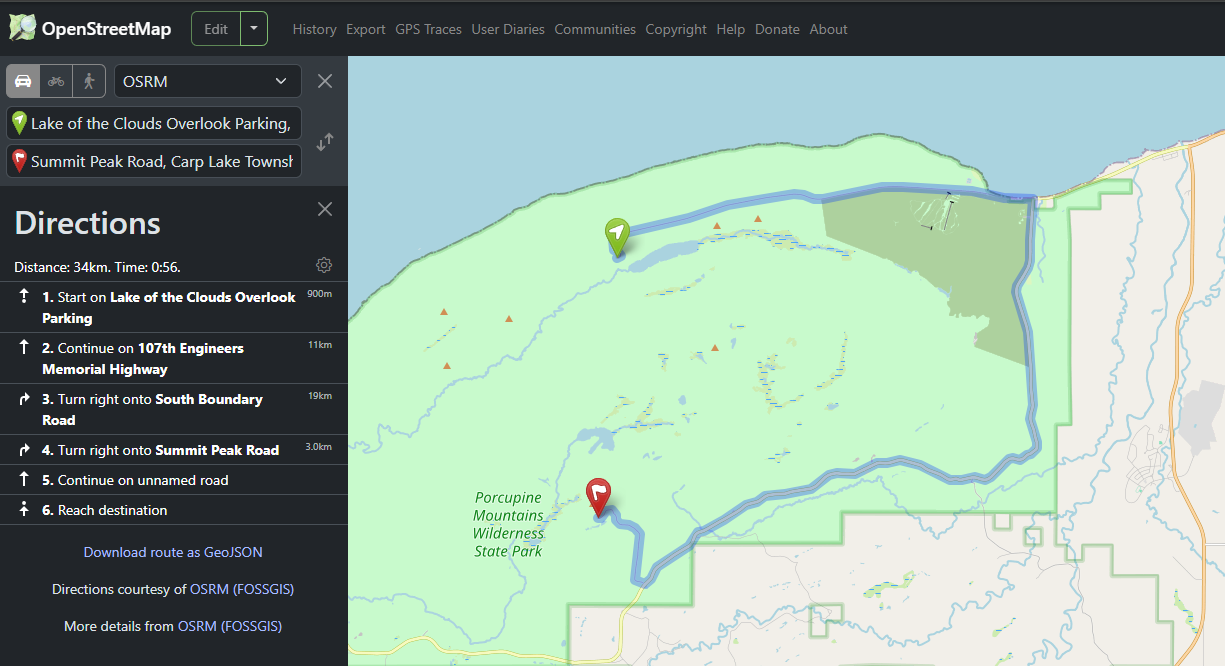}
    \caption{Map View of the chosen Hilly Terrain Route.}
    \label{fig:hilly_terrain_route}
\end{figure}

\begin{figure}[htbp]
    \centering
     \includegraphics[width=\linewidth]{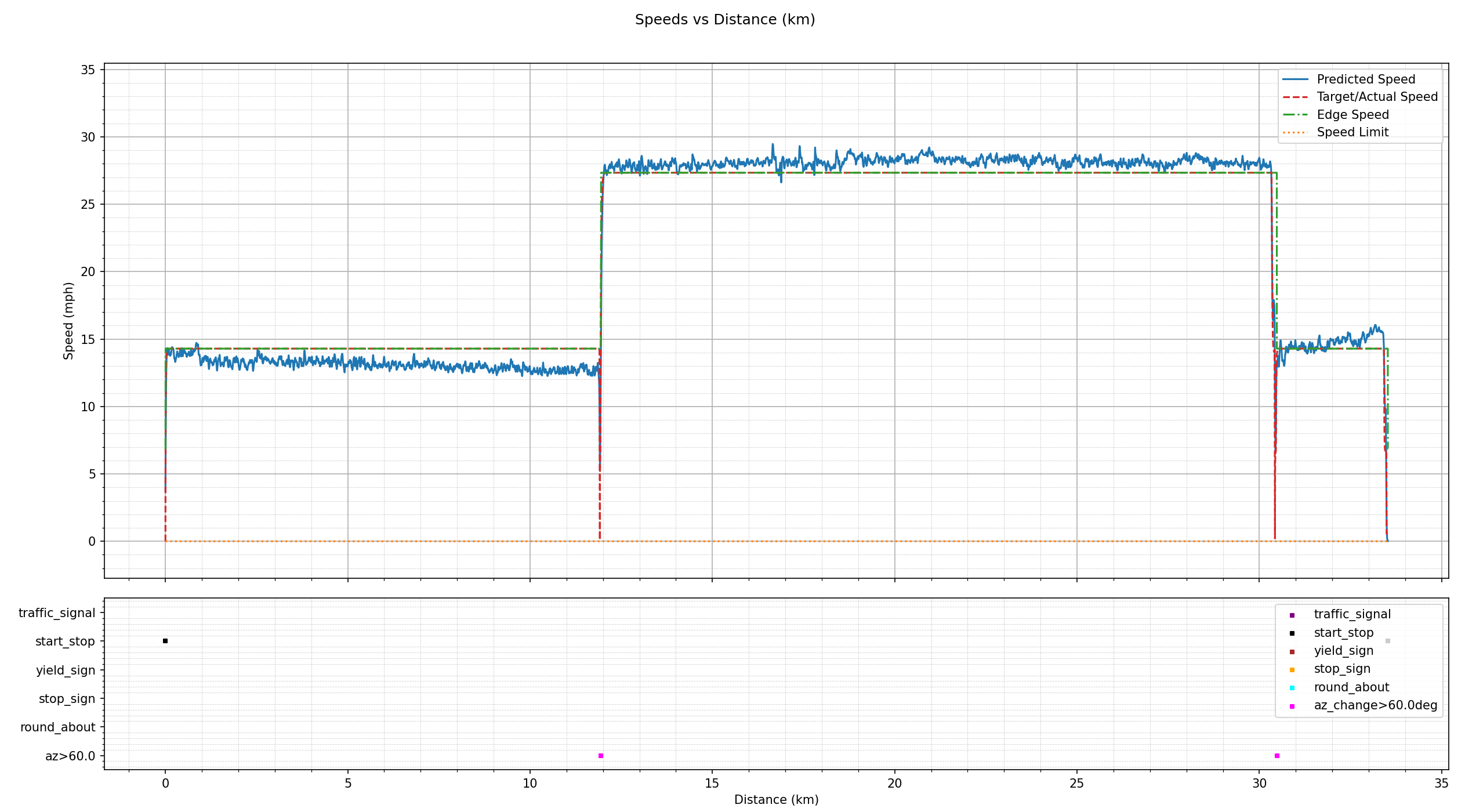}
    \caption{Predicted Velocity for the chosen  Hilly Terrain Route.}
    \label{fig:hilly_terrain_vel}
\end{figure}

\begin{figure}[htbp]
    \centering
     \includegraphics[width=\linewidth]{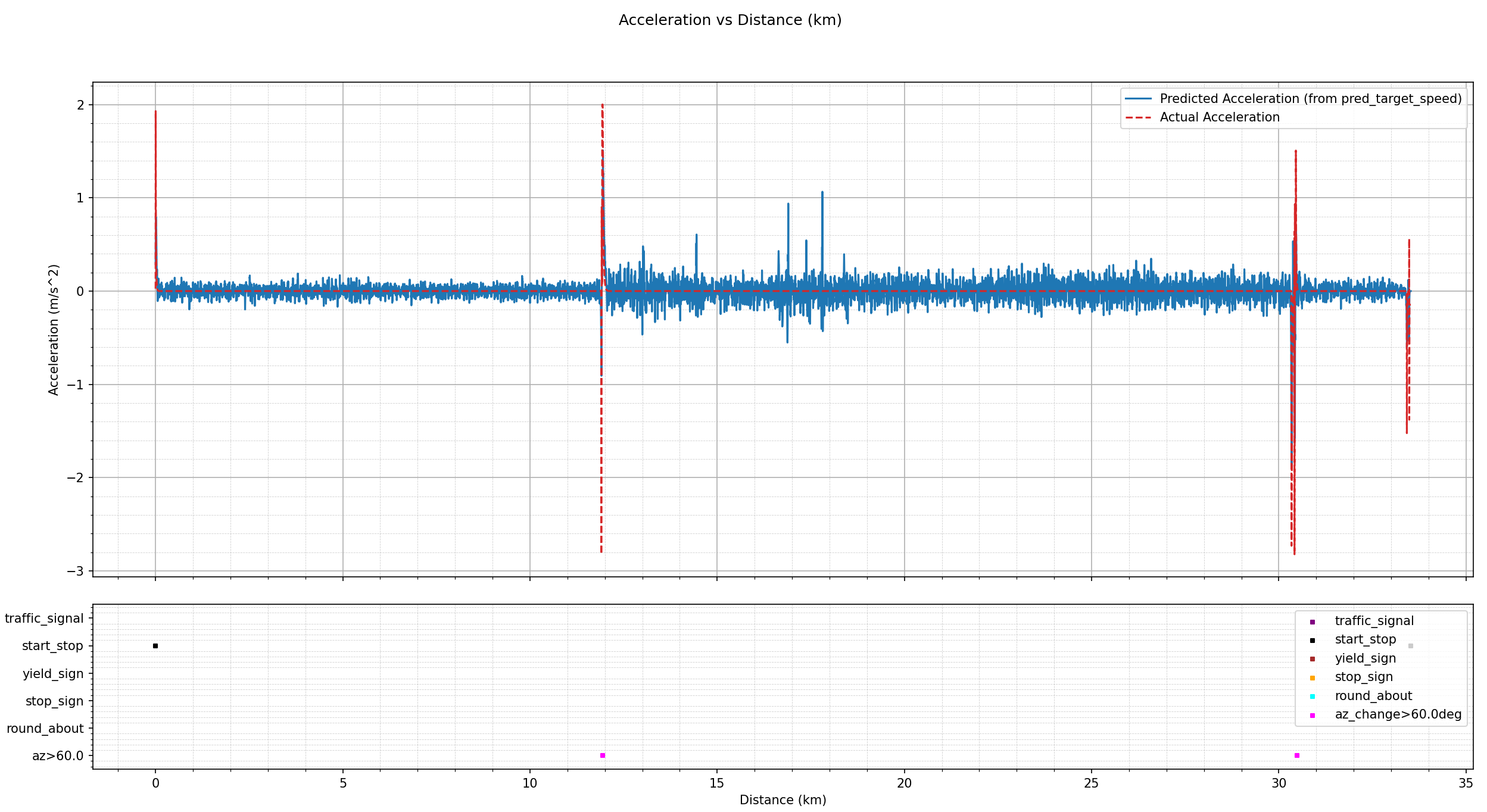}
    \caption{Predicted Acceleration for the chosen  Hilly Terrain Route.}
    \label{fig:hilly_terrain_acc}
\end{figure}

A hilly route in the Porcupine Mountains of Michigan shown in Fig.~\ref{fig:hilly_terrain_route} used for this test case includes an initial descent followed by a sustained climb. The driver-specific predicted velocity in Fig.~\ref{fig:hilly_terrain_vel} shows the subtle influence of elevation changes on vehicle speed. The corresponding predicted acceleration profile is shown in Fig.~\ref{fig:hilly_terrain_acc}.

\begin{figure}[htbp]
    \centering
     \includegraphics[width=\linewidth]{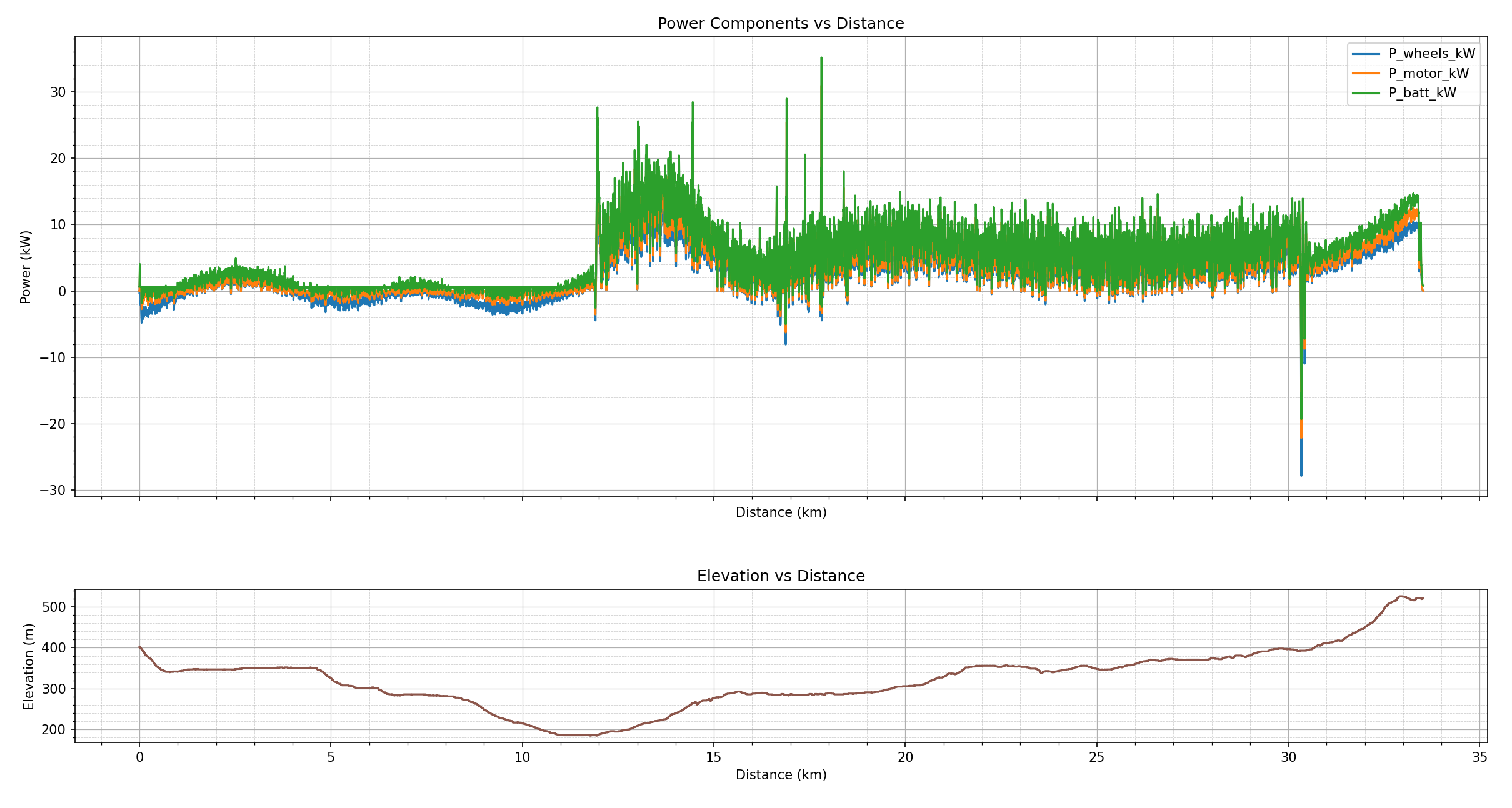}
    \caption{Predicted power demand for the chosen  Hilly Terrain Route.}
    \label{fig:hilly_terrain_power}
\end{figure}

\begin{figure}[htbp]
    \centering
     \includegraphics[width=\linewidth]{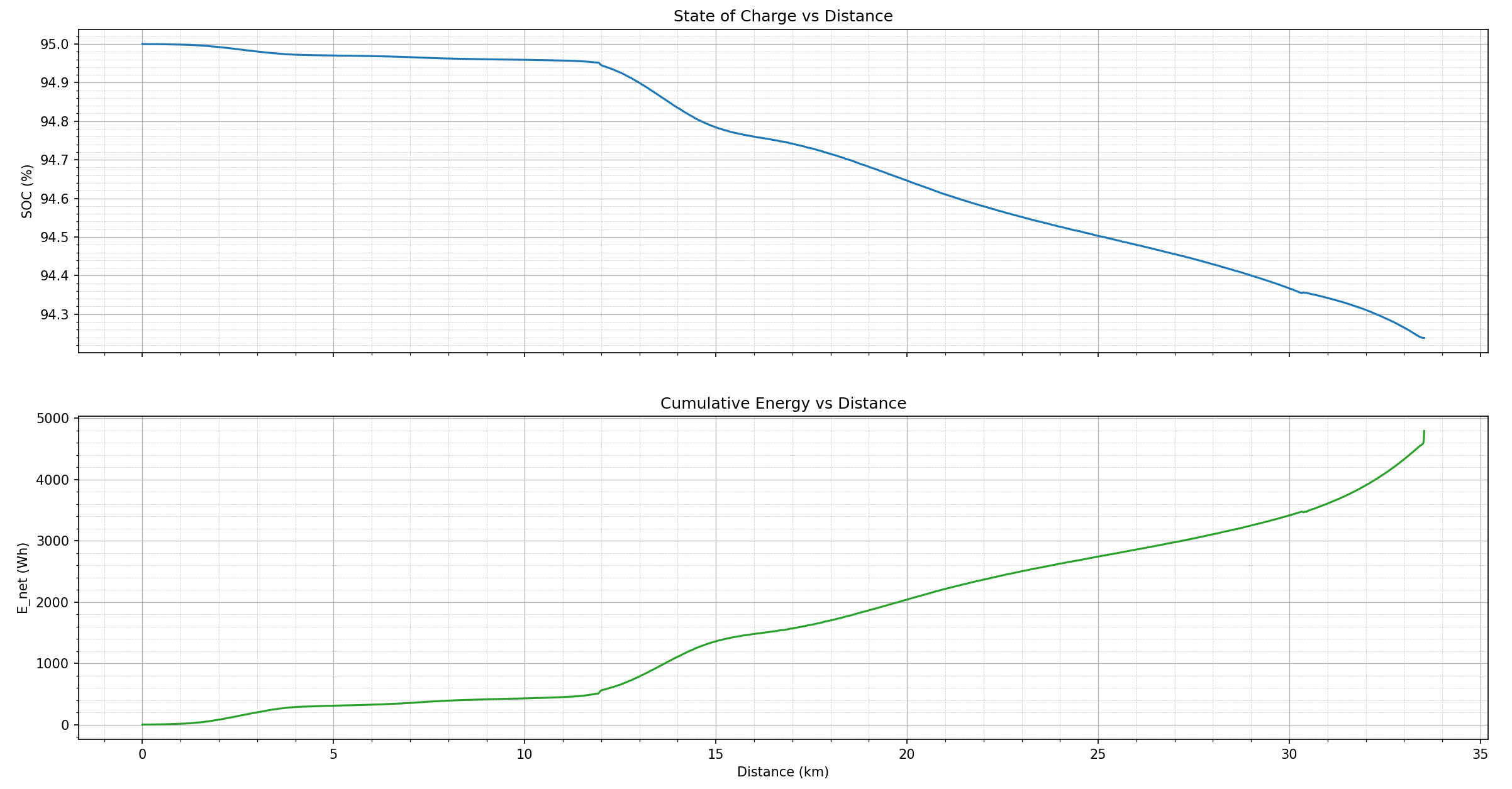}
    \caption{Predicted energy Consumption for the chosen Hilly Terrain Route.}
    \label{fig:hilly_terrain_soc}
\end{figure}

Power and SOC profiles shown in Fig.~\ref{fig:hilly_terrain_power} and Fig.~\ref{fig:hilly_terrain_soc}, respectively, align with physical expectations: downhill sections exhibit predominantly negative power due to continuous regeneration, while uphill sections show consistently positive power and accelerated SOC depletion. Consequently, the downhill portion exhibits slow SOC depletion, while the uphill portion shows a much faster rate of energy consumption and SOC decline.


\section{Conclusion}

This work presented an integrated framework for personalized BEV range estimation that combines map-derived contextual features, driver-specific velocity prediction using an LSTM model, and a physics-based energy and SOC estimation module. The end-to-end pipeline---comprising the Route Processor, Velocity Profile Generator, PID Controller, Driver LSTM Model, and BEV Energy Consumption Model---produced realistic velocity, power, and SOC trajectories across urban, freeway, and hilly routes. System-level evaluations demonstrated that the LSTM model captures key behavioral patterns, including deceleration near intersections, speed-limit tracking on freeways, and grade-dependent responses on hilly terrain. The resulting energy and SOC profiles aligned with expected physical behavior, such as regeneration during downhill segments and increased power demand during hill climbs.

The results highlight the effectiveness of combining learned driver behavior with physics-based modeling for context-aware BEV performance prediction. The modular architecture also supports extensibility and experimentation.

Future improvements include enhancing stop-event behavior through longer look-ahead context and richer training data, exploring attention-based or transformer architectures for variable-length temporal dependencies, and refining the energy model with detailed component-level efficiency maps. Broader evaluation across additional drivers, routes, and environmental conditions would further strengthen generalization.

Overall, this framework provides a foundation for accurate, personalized, and route-aware BEV range estimation, enabling applications in eco-routing, energy-aware navigation, and driver-adaptive energy management.

\end{document}